\documentclass[%
 aip,
 amsmath,amssymb,
 reprint,%
]{revtex4-1}
\usepackage{graphicx}

\usepackage{dcolumn}
\usepackage{bm}

\usepackage[utf8]{inputenc}
\usepackage[T1]{fontenc}
\usepackage{mathptmx}
\usepackage{gensymb}
\usepackage{caption}
\usepackage{subcaption}

\draft 

\begin{document}


\title{On-chip Integration of Si/SiGe-based Quantum Dots and Switched-capacitor Circuits} 



\author{Y. Xu}
\affiliation{QuTech and Kavli Institute of Nanoscience, Delft University of Technology, Delft, the Netherlands}

\author{F. K. Unseld}
\affiliation{QuTech and Kavli Institute of Nanoscience, Delft University of Technology, Delft, the Netherlands}

\author{A. Corna}
\affiliation{QuTech and Kavli Institute of Nanoscience, Delft University of Technology, Delft, the Netherlands}

\author{A. M. J. Zwerver}
\affiliation{QuTech and Kavli Institute of Nanoscience, Delft University of Technology, Delft, the Netherlands}

\author{A. Sammak}
\affiliation{QuTech and Netherlands Organization for Applied Scientific Research (TNO), Delft, the Netherlands}

\author{D. Brousse}
\affiliation{QuTech and Netherlands Organization for Applied Scientific Research (TNO), Delft, the Netherlands}

\author{N. Samkharadze}
\affiliation{QuTech and Netherlands Organization for Applied Scientific Research (TNO), Delft, the Netherlands}

\author{S. V. Amitonov}
\affiliation{QuTech and Kavli Institute of Nanoscience, Delft University of Technology, Delft, the Netherlands}

\author{M. Veldhorst}
\affiliation{QuTech and Kavli Institute of Nanoscience, Delft University of Technology, Delft, the Netherlands}

\author{G. Scappucci}
\affiliation{QuTech and Kavli Institute of Nanoscience, Delft University of Technology, Delft, the Netherlands}

\author{R. Ishihara}
\affiliation{QuTech and Kavli Institute of Nanoscience, Delft University of Technology, Delft, the Netherlands}

\author{L. M. K. Vandersypen}
\affiliation{QuTech and Kavli Institute of Nanoscience, Delft University of Technology, Delft, the Netherlands}

\date{\today}

\begin{abstract}
Solid-state qubits integrated on semiconductor substrates currently require at least one wire from every qubit to the control electronics, leading to a so-called wiring bottleneck for scaling. Demultiplexing via on-chip circuitry offers an effective strategy to overcome this bottleneck.
In the case of gate-defined quantum dot arrays, specific static voltages need to be applied to many gates simultaneously to realize electron confinement. When a charge-locking structure is placed between the quantum device and the demultiplexer, the voltage can be maintained locally.
In this study, we implement a switched-capacitor circuit for charge-locking and use it to float the plunger gate of a single quantum dot. Parallel plate capacitors, transistors and quantum dot devices are monolithically fabricated on a Si/SiGe-based substrate to avoid complex off-chip routing. We experimentally study the effects of the capacitor and transistor size on the voltage accuracy of the floating node. Furthermore, we demonstrate that the electrochemical potential of the quantum dot can follow a 100 Hz pulse signal while the dot is partially floating, which is essential for applying this strategy in qubit experiments.       
\end{abstract}

\pacs{}

\maketitle 

Silicon spin qubits based on gate-defined quantum dots have recently been realized on Si/SiGe substrates with high fidelity~\cite{Maune2012, Erika2014, Zajac2018, Watson2018, Yoneda2018, Xiao2019}. Given their compatibility with current semiconductor fabrication techniques and potential for easy integration with classical electronics, these quantum dots are considered to be a promising basis for quantum computers~\cite{Intel_IEDM2018,Eriksson2013,Divincenzo1988}. Nevertheless, a fault-tolerant quantum computer requires millions of qubits~\cite{Surfacecode2012}. Even if quantum dots are designed to be identical, the required gate bias voltage still differs among the dots due to non-uniformities in the substrate and variations during the fabrication process. 
In a standard operating mode, two gates are used to control each quantum dot on average and each gate is connected to a separate room temperature digital-to-analog converter (DAC) through the bond wires from the chip to the sample carrier and the dilution refrigerator wiring. However, 
this linear approach clearly poses a bottleneck to scaling up the number of qubits. By comparison, today's classical processor chips have only about 2000 contact pins, while billions of transistors can be integrated and operated on a single chip. This large ratio between active components and pins is described by Rent's rule and is made possible by implementing shared control methods\cite{franke2018}. In order to operate the millions of qubits for practical quantum computation, similar methods will therefore have to be implemented in quantum integrated circuits.  

Inspired by the classical dynamic random-access memory (DRAM) matrix that uses word lines and bit lines to address a large number of storage cells~\cite{DRAM2007}, proposals for controlling spin qubits using word and bit lines exist~\cite{Hollenberg2015,Lieven2017,Menno2017,Ruoyu2018,Jelmer2019}. Another concept that can be borrowed from DRAM is charge-locking, which, when combined with demultiplexers, allows to significantly reduce the number lines going off-chip~\cite{Puddy2015,Lieven2017,Menno2017,Reilly2019}. In DRAM, the stored voltage encodes a “0” or “1”, according to a threshold. In contrast, the voltage maintained on a quantum dot gate needs to be a precise analog value.  The required precision of such a stored voltage ranges from 1 $\mu$V to 1 mV, depending on the gate function and coupling of the gate to the dot potential~\cite{Lieven2017,Jelmer2019}. Charge-locking is thus used in the form of a sample-and-hold circuit, when the input line is electrically detached, the gate of the quantum dot is floating and the voltage maintains there for a certain period. 
Although the primary role for DC gates of quantum dots is to achieve electron confinement, additional voltage pulses must be applied to these gates for qubit experiments. For example, in a commonly used single-shot readout method to determine the state of an electron spin, a few kHz signal is applied to the gate to load, read and empty a quantum dot~\cite{Elzerman2004,Morello2010}. When a switched-capacitor (SC) circuit is integrated with these gates as an interface, the extra transistor or capacitor should not affect the voltage pulses arriving at the quantum dot gates. 
Prototypes have been made with on-chip or off-chip integrated floating gate circuits and GaAs quantum dots~\cite{Puddy2015,Reilly2019}. For silicon-based quantum dots, switching circuits have been integrated with quantum devices on-chip~\cite{Cryogenic_on-chip_mux2013,Integration_of_on-chip_FET2013}. In addition, charge-storage devices and quantum devices have been fabricated using the same CMOS process and connected through wire bonds~\cite{Schaal2018, Schaal2019}. However, in silicon a fully on-chip integrated solution, without the need for wire bonds, is still waiting to be achieved.

In this study, we integrate a switched-capacitor circuit containing an n-type transistor and a holding capacitor with a single quantum dot on a Si/SiGe-based substrate. We analyze the parameters that affect the variability of the floating gate voltage and experimentally study the impact of the size of the capacitor and the transistor. In addition, we apply a pulsed voltage to one of the quantum dot gates while floating another gate, as a relevant test for qubit measurements.
\begin{figure}
\centering
\begin{subfigure}[t]{0.33\linewidth}
\setlength{\abovecaptionskip}{-0.5mm}
\caption*{(a)}
\includegraphics[width=\linewidth]{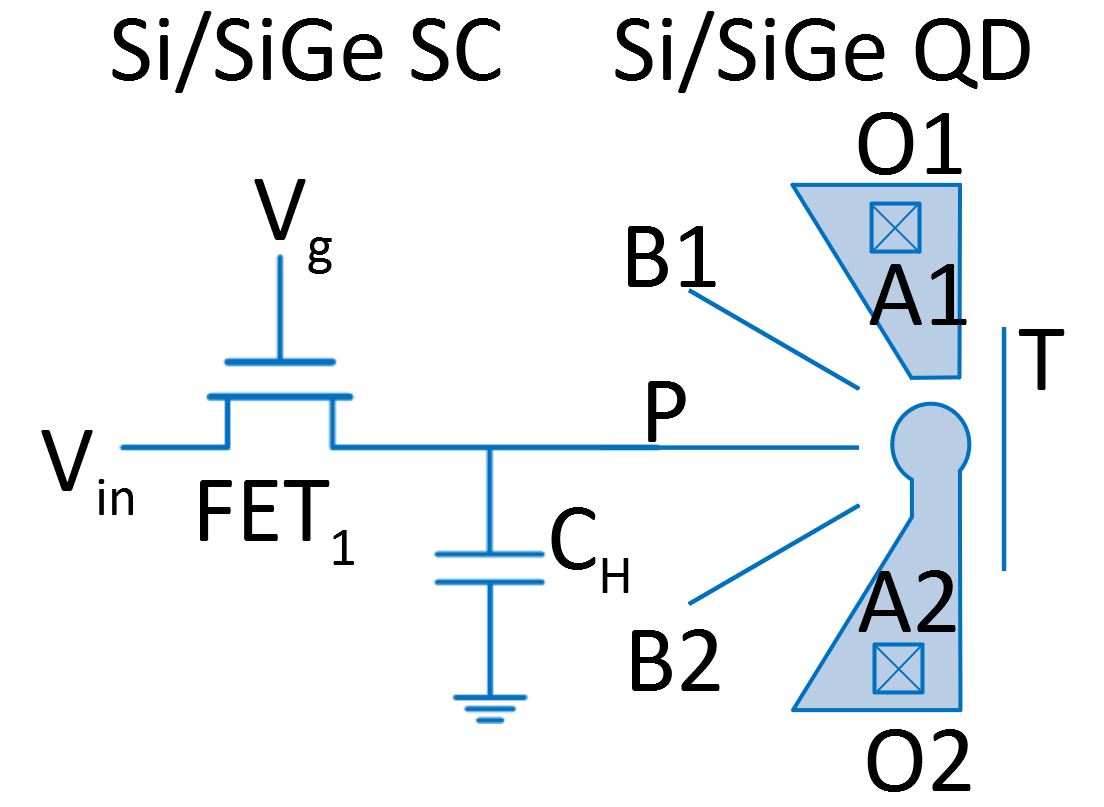}
\end{subfigure}
\begin{subfigure}[t]{0.66\linewidth}
\setlength{\abovecaptionskip}{-0.4mm}
\caption*{(c)}
\includegraphics[width=\linewidth]{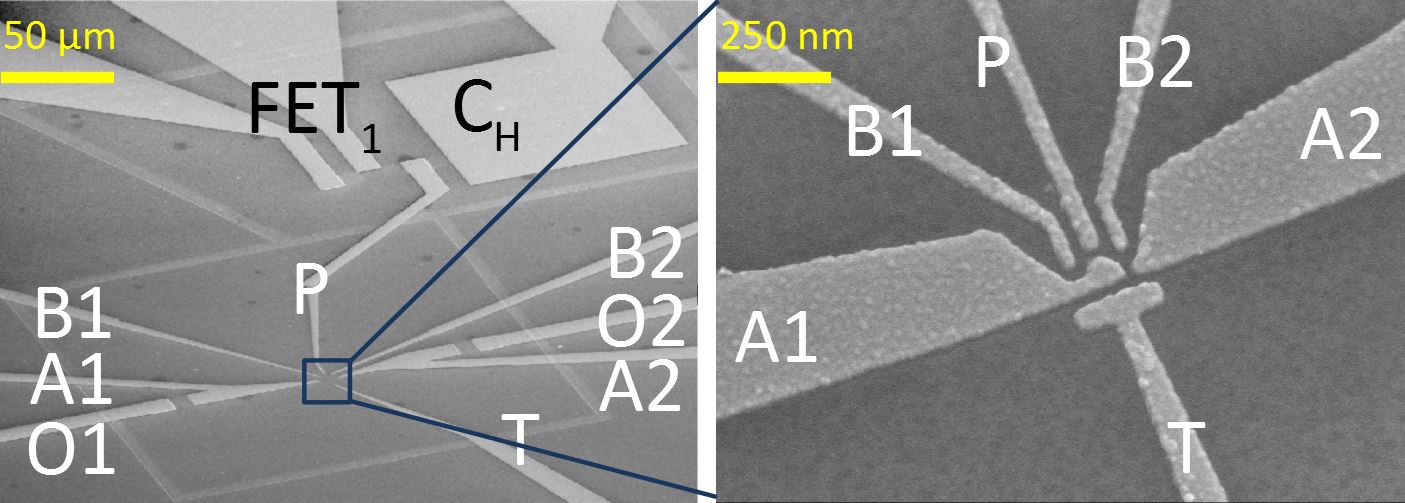}
\end{subfigure}
\begin{subfigure}{\linewidth}
\setlength{\abovecaptionskip}{-0.5mm}
\caption*{(b)}
\includegraphics[width=\linewidth]{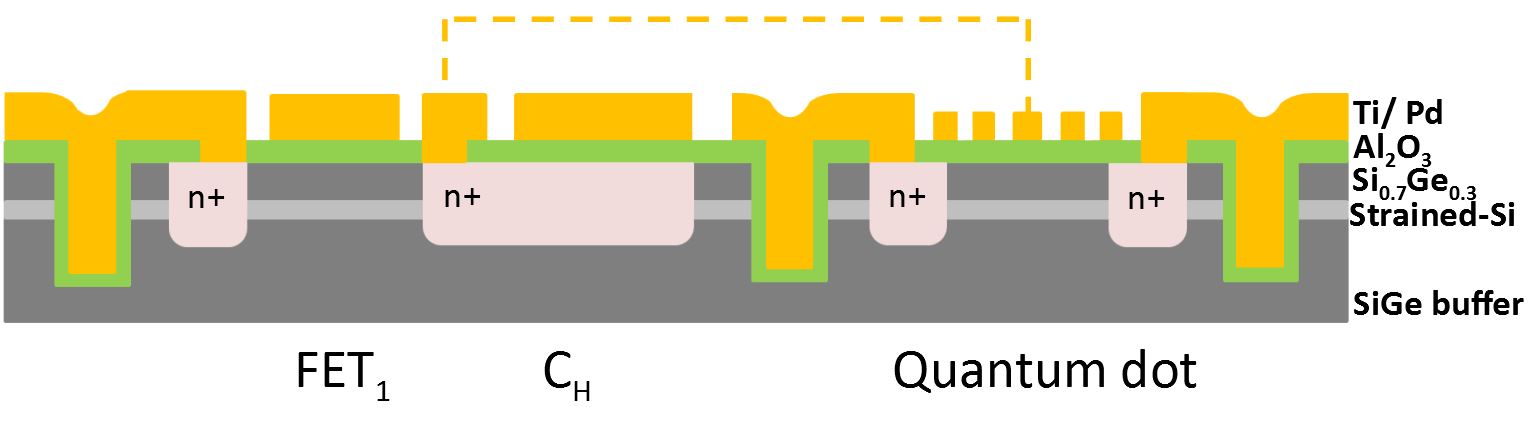}
\end{subfigure}
\caption{\label{fig:device}(a) Device schematic showing the switched-capacitor circuit connected to gate $P$ of a single quantum dot. (b) Cross-section of the device. All elements, the transistors, capacitors, and quantum dots are integrated on the same Si/SiGe based substrate and share the same aluminum oxide layer as dielectric. (c) SEM image of device $C$. The relevant device dimensions are listed in Table \ref{tab:table1}.}
\end{figure}

The impact of the design parameters on the floating node voltage accuracy as described in the literature provides guidance for our choices of device dimensions (see Fig.\ref{fig:device}(a)). We first review two mechanisms that lead to a random error in the floating node voltage, and next describe two mechanisms that  produce a systematic error. As we shall see, in general a larger holding capacitance reduces these errors, but it also increases the footprint and power dissipation, both of which can limit scalability as well~\cite{franke2018}. 

Fundamentally, the voltage resolution $\varDelta V$ of a floating node is limited by the electron charge, $e$, as 
\begin{equation}
\label{eq:single_electron}
\varDelta V = e/C_H \;,
\end{equation}
where $C_H$ is the total capacitance of the floating node to ground. It is dominated by the holding capacitor in our case. In order to keep $\varDelta V$ below 1 $\mu$V, $C_H$ should be larger than 160 fF.

Next, thermal noise is present due the transistor channel resistance when the transistor (\textit{FET$_1$}) in a switched-capacitor circuit is switched on. The random thermal noise voltage is maintained on the holding capacitor after switching off the transistor. The root-mean-square (RMS) noise voltage on the capacitor is calculated as~\cite{noise1979} 
\begin{equation}
\label{eq:noise}
V_{n}^{rms} = \sqrt{k_BT{/}C_H} \;.
\end{equation}
For instance, to obtain a noise level below 1 $\mu$V at a temperature of 10 mK, the holding capacitance must exceed 138 fF.

A first systematic offset in the floating node voltage is caused by channel charge injection. This effect refers to the charges that get redistributed to the drain and source upon switching a transistor off\cite{Analog_CMOS_IC2001,Analog_IC_design1997}. Under the assumption that charges split equally to the source and drain, the error $\varDelta V_c$ in the stored voltage on the floating node can be expressed as
\begin{equation}
\label{eq:channel_charge_injection}
\varDelta V_c = \frac{C_{channel}(V_g^{ON}-V_{in}-V_{th})}{2C_H} \;,
\end{equation}
with $C_{channel}$ the capacitance between the transistor gate and channel, $V_g^{ON}$ the “on” voltage on the gate of \textit{FET$_1$}, $V_{in}$ the input voltage indicated in Fig.1(a), and the threshold voltage $V_{th}$ is the voltage difference between gate and source/drain at which charges begin to accumulate in the channel. For instance, when $V_g^{ON}$  is set 0.1 V higher than $V_{in}+V_{th}$, the holding capacitance needs to be 50 times larger than the transistor channel capacitance to keep $\varDelta V_c$ below 1 mV. 

Another factor that introduces systematic offsets in the maintained voltage is the parasitic capacitance from the transistor gate to the floating node. In series with the holding capacitance, it shifts the voltage on the floating node by an amount that depends on the voltage on the gate of \textit{FET$_1$}, given by~\cite{Analog_CMOS_IC2001,Analog_IC_design1997}
\begin{equation}
\label{eq:paracitic_Cgs}
\varDelta V_p = \varDelta V_g \frac{C_{gs}}{C_{gs}+C_H} \;,
\end{equation}
where $\varDelta V_g$ is the switching range used to turn the transistor on and off ($\varDelta V_g = V_g^{ON}-V_g^{OFF}$). 
Taking $\varDelta V_g$  as 1 V, the ratio of $C_H$ to $C_{gs}$ should exceed 1000 to keep $\varDelta V_p$ below 1 mV.

Importantly, different from the random variations in the floating gate voltage, the systematic shifts can be accounted for in the calibration phase, hence they do not impose strict requirements on $C_H$. 

Turning now to power dissipation, the heat generated from the on-resistance of \textit{FET$_1$} and the parasitic resistance on the leakage path can be expressed as   
\begin{equation}
\label{eq:power1}
P_1 = C_Hf_g{(V_1-V_2)}^2 \;,
\end{equation}
where $V_1$  and $V_2$ are the high and low voltages on the holding capacitor during operation, and $f_g$ is the switching frequency of the transistor. Note that $P_1$ is  proportional to the holding capacitance. If we refresh the floating node to compensate a 1 mV drop with a 1 Hz frequency, the power dissipation of a single cell is $10^{-18}$ W when the holding capacitor is 1 pF. This is orders of magnitude smaller than the heat dissipated upon switching in the resistance in the line between the pulse generator and the transistor gate, which is given by   
\begin{equation}
\label{eq:power2}
P_2 = \frac{1}{2}C_{channel}f_g{(V_{g}^{ON}-V_{g}^{OFF})}^2 \;.
\end{equation}
For a transistor with 0.01 pF channel capacitance and 1 V switching range, the power dissipated on the signal line to its gate is $5\times10^{-13}$ W. Even if we assume that this power is entirely dissipated on-chip, it would still allow $2\times10^{8}$ floating gate voltages to be maintained assuming 100 $\mu$W available cooling power at the chosen operating temperature.

Making the transistors smaller reduces $C_{channel}$ and $C_{gs}$, which reduces switching power dissipation as well as the systematic shifts in the floating gate voltage. However, secondary effects appear when the device is scaled down. For instance, when the channel width is below 1 $\mu$m, the threshold voltage $V_{th}$ increases due to the narrow-channel effect~\cite{Semiconductor_physics2012, High_Vth2016}. Then a higher gate voltage is required to turn on the transistor which is more likely to cause hysteresis and breakdown.

\begin{table}
\caption{\label{tab:table1} Device dimensions and voltage variations on the floating node between static and floating mode tests}
\begin{ruledtabular}
\begin{tabular}{cccc}

               & Device $A$ & Device $B$ & Device $C$
\\ \hline
$C_H$ size ($\mu$m $\times$ $\mu$m)   & 15$\times$15 & 15$\times$15  & 100$\times$100 \\
$C_H$ (pF)   & 0.697 & 0.697  & 30.98 \\
\textit{FET$_1$} size ($\mu$m $\times$ $\mu$m)   & 10$\times$10 & 10$\times$1  & 10$\times$1 \\
$C_{channel}$ (pF) & 0.171  & 0.022  & 0.022 \\ 
$C_{gs}$ (fF) & 30  & 3.9  & 3.9 \\
Expected $\varDelta V_p$ (mV) & 53.65  & 2.67  & 0.08 \\ 
Expected $\varDelta V_c$ (mV) & 9.3-26.4  & 2.9-3.5  & 0.04-0.08 \\
$\sqrt{kT{/}C_H}$ ($\mu$V) & 0.44 & 0.44 & 0.06 \\
$e/C_H$ ($\mu$V)  & 0.23  & 0.23  & 0.0052 \\ 
Measured shift (mV) & 44-48  & 2.8-5.4  & 0.5-1 \\
\end{tabular}
\end{ruledtabular}
\end{table}

\begin{figure}
\includegraphics[width=\linewidth]{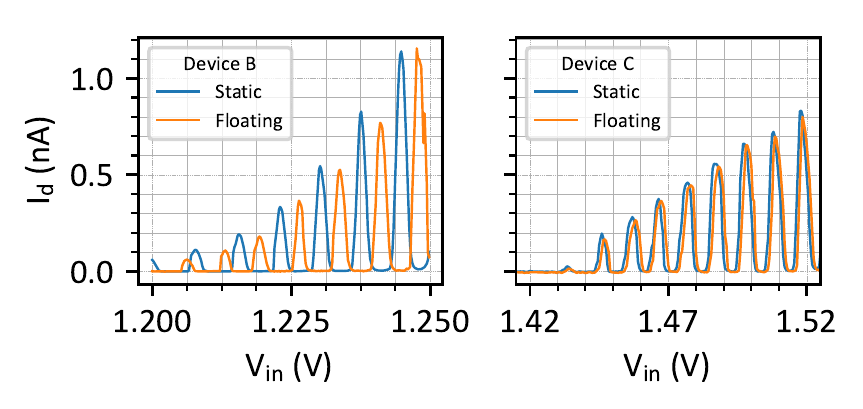}
\centering
\vspace{-0.7cm}%
\caption{\label{fig:static_float} Current through the quantum dot as a function of $V_{in}$, with device $B$ (left panel) and $C$ (right panel) operated in static and floating mode. Patterns of Coulomb peaks are consistent between the blue (static mode) and orange (floating mode) traces. The voltage shift for device $C$ is less than that for device $B$ due to its larger holding capacitance.}
\end{figure}
Based on the above considerations, we made three devices with the same quantum dot design but different transistor and holding capacitor sizes and compared their voltage variations on the floating node both theoretically and experimentally. The device dimensions are listed in Table~\ref{tab:table1} (see also Appendix A). One practical consideration for the length of the transistor is the lateral diffusion of the implantation region, which is estimated to be 0.4 $\mu$m in this process. The transistor channel length is chosen to be 10 $\mu$m to reduce the effect of lateral diffusion on the actual length of the channel, though fundamentally the length could be much reduced.

For all devices, the quantum dot, transistor, and capacitor were integrated on a Si$_{0.7}$Ge$_{0.3}$/strained-Si/buffered-SiGe heterostructure substrate~\cite{Nodar2018}. For the quantum dot, we used a single patterned metal layer to define the potential landscape that confines electrons. The top plate of the capacitor is formed by a metal gate and the bottom plate by a heavily implanted region in the semiconductor, with a dielectric separating the plates. The transistor is a field-effect transistor with the buried quantum well acting as the channel (see Fig.\ref{fig:device}(b)).

The fabrication process began with the definition of markers followed by phosphorus ion implantation to create reservoirs for the quantum dots, the source and drain of the transistors, and the electrodes for the capacitors. This was followed by rapid thermal annealing at 700 $\degree$C to activate the dopants. Trenches of 100 nm deep and 10 $\mu$m wide were subsequently etched into the Si/SiGe substrate to isolate the devices from each other. A 20 nm Al$_2$O$_3$ layer was then grown via atomic layer deposition to form the gate oxide for both quantum dots and transistors. Finally, we used electron beam lithography and lift-off to pattern an electron-beam evaporated 5/15 nm Ti/Pd stack to define the quantum dot gates, followed by patterning of a 5/195 nm Ti/Pd film for the transistor gate, the top electrode of the capacitor and the leads and pads of the quantum dot. Fig.\ref{fig:device}(c) shows the SEM image of one of the final devices (Device $C$ in Table \ref{tab:table1}). Details regarding the separate characterization of the single quantum dot and the transistor are described in Appendix B. 

We mounted the device in a dilution refrigerator operating at a base temperature below 10 mK and at zero magnetic field. All current measurements through the quantum dot are performed with a 100 $\mu$V source-drain bias applied across the quantum dot.

As a reference, the device was first tested in static mode with gate $P$ not floating (blue traces in Fig.\ref{fig:static_float}). The current through the quantum dot was experimentally measured while the transistor was conducting. In the floating mode tests, we first turned on \textit{FET$_1$} to charge the holding capacitor and then turned it off. After 10 ms, we measured the current through the quantum dot while gate $P$ was floating. As shown in Fig.\ref{fig:static_float} (and Fig. S1 for device $A$), the patterns of the Coulomb peaks measured in floating mode were consistent with those measured in static mode, but shifted in $V_{in}$. Table~\ref{tab:table1} summarizes the expected and measured voltage shifts. The measured dependence of gate voltage shift versus dimensions matches the predicted trend very well, with device $A$ showing the largest shifts and device $C$ the smallest shifts.

Nevertheless, there are still variations between the measured and expected voltage shifts. We here discuss this difference for device $B$. First we note that the average voltage decay rate for device $B$ in the first 40 seconds after opening the transistor was approximately 2.8 $\mu$V/s (see Appendix C). This very low leakage rate compared to commercial DRAM is possible owing to the low operating temperature. The voltage shift on the floating gate due to leakage through the holding capacitor is thus negligible during the 10 ms interval between the moment the transistor is opened and the time of measurement. We will therefore compare the measured voltage shifts to those expected based on Eqs.~\ref{eq:single_electron}-\ref{eq:paracitic_Cgs}. Fig. S6(b) in Appendix F shows the measured voltage shifts for the consecutive Coulomb peaks for device $B$. The overall trend of peak shift versus $V_{in}$ matches well with Eq.~\ref{eq:channel_charge_injection}, expressing charge injection from the channel, for the five Coulomb peaks at the highest $V_{in}$ (the leftmost peak is shifted more than expected). The additional overall systematic shift is smaller than that expected based on Eq.~\ref{eq:paracitic_Cgs} (the transistor gate voltage coupling in through the parasitic gate-source capacitance), possibly in part due to deviations in the estimated dielectric thickness or constant.  In addition, the individual shifts do fluctuate around the overall linear trend by about $\pm 0.2$ mV (see Appendix E). By comparison, the random shifts expected from charge quantization and thermal noise (Eqs.~\ref{eq:single_electron}-\ref{eq:noise}) are below 1 $\mu$V. However, the measured voltage fluctuations match well with the measured 1/$f$ noise caused by background charge fluctuations modulating the dot potential. The Coulomb peak measurement took a few minutes to complete and the 1/$f$ noise amplitude at 0.01 Hz is indeed of order $0.2$ mV/$\sqrt{\mbox{Hz}}$ (see Appendix D).

For qubit operation and readout, gate voltage pulses must be applied to one or more of the quantum dot gates. We now test the compatibility of applying such pulses with a switched-capacitor circuit present and operated in floating mode. In principle the voltage pulses can be applied either to a floating gate (e.g. via the holding capacitor) or to another gate. Either way, the question is to what extent the presence of the capacitor and transistor that form the SC circuit distorts the waveform. 

\begin{figure}
\includegraphics[width=\linewidth]{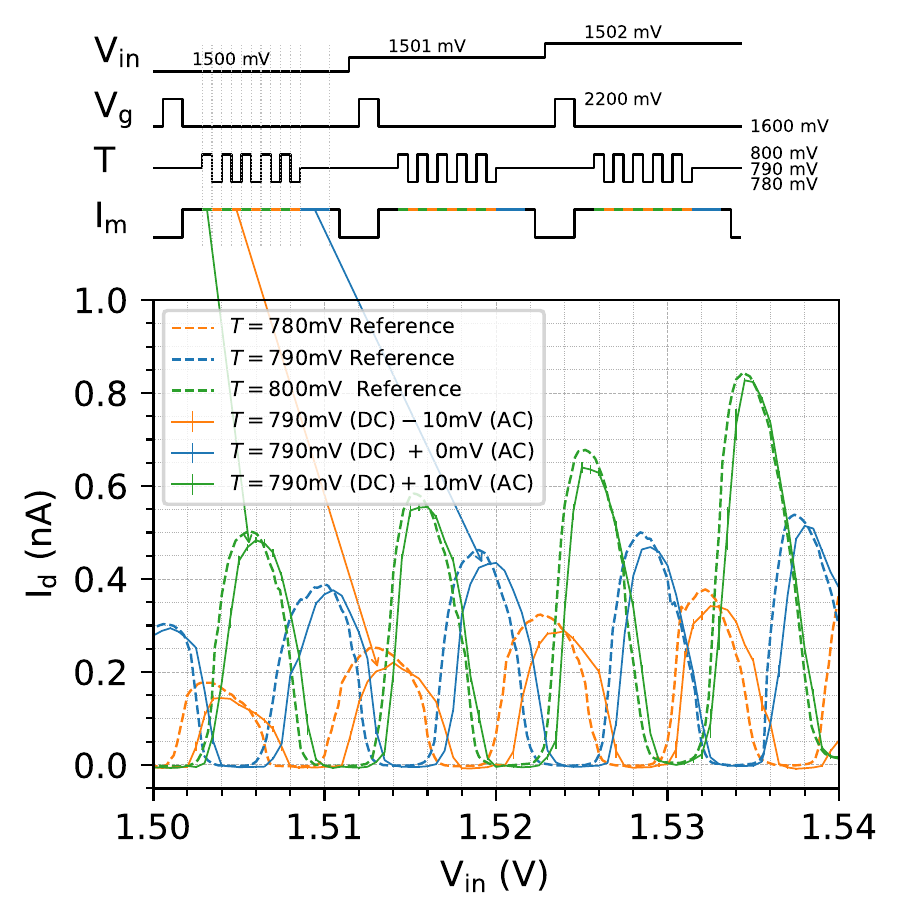}
\vspace{-0.5cm}%
\caption{\label{fig:Figure3}  Gate $P$ of device $C$ was floated for 200 ms for each value of $V_{in}$. A 140 mV$_{pp}$, 100 Hz square wave was applied through a 23 dB attenuator to gate $T$ in the first 100 ms, adding to a DC bias voltage of 790 mV on the same gate. The current through the quantum dot was measured throughout these operations. Data corresponding to the high and low level of the voltage pulse (T=800 mV and T=780mV), and to the second 100 ms without pulses (T=790 mV), are plotted separately as a function of $V_{in}$ (solid lines, the error bars indicate the standard deviation of each data point). The Coulomb peaks correspond very well to the static reference measurements, obtained with 780  mV, 790 mV and 800 mV DC voltages directly applied on gate $T$ while gate $P$ was not floating  (dashed lines).}
\end{figure}

Here we perform a preliminary test for voltage pulses applied to a gate that is not floating. 
Limited by the 1 kHz sampling rate of the current measurement, we provided a 100 Hz square wave to gate $T$ (indicated in Fig. \ref{fig:device}(a)) of device $C$, and check whether the electrochemical potential of the quantum dot is able to follow the signal while gate $P$ is floating. 
The sequence of the experiment is depicted in Fig.\ref{fig:Figure3}. The input voltage $V_{in}$ is stepped through a range that covers several Coulomb peaks. For each $V_{in}$, we first floated gate $P$ by setting $V_g$ from “high” to “low”. Then a 100 Hz, $\pm$ 10 mV pulse signal was applied to gate $T$ through a bias tee during 100 ms, adding to a 790 mV DC bias voltage, while we continuously measured the current flowing through the quantum dot. The current through the dot corresponding to the two stages of the voltage pulse, as well as the current during a subsequent time interval without gate voltages pulses, were extracted separately and compared to the static mode measurement results as shown in Fig.\ref{fig:Figure3}. The Coulomb peak patterns were consistent with the respective reference measurements. The 0.6-1.0 mV voltage shift of the center peaks (blue solid versus dotted traces) is in agreement with the expected shift from channel charge injection and parasitic capacitance of the transistor upon switching off. Furthermore, the peaks obtained while applying a 100 Hz square pulse overlap closely with their expected positions, see the green and orange solid and dotted lines. The 0.6 mV larger average shift for the orange versus the green solid lines indicates that the square pulse amplitude at the gate is slightly larger than the intended $\pm 10$ mV, which is explained by a deviation (within the specified tolerance) of values of the attenuators placed in the transmission line connected to gate $T$. These results show that the voltage pulses on gate $T$ were not affected by the switched-capacitor circuit and by floating gate $P$. Based on electric circuit simulations (see Appendix G) that include the various capacitors discussed in the text, we expect the large capacitor that stores the floating node voltage (on gate $P$) not to impact the modulation of the dot potential in response to a pulse on gate $T$ until at least 20 GHz.

In summary, in this study we demonstrated that a switched-capacitor circuit placed between a quantum dot and demultiplexer can function as a local voltage source. The effect of channel charge injection and gate-source capacitive coupling introduce a systematic offset on the sampled voltage, which can be reduced if desired by using a larger holding capacitor and a smaller transistor. In the present measurements, random offsets in the stored voltage are dominated by 1/$f$ noise in the dot potential. Finally, we show that floating a quantum dot gate does not impact the effect of (slow) voltage pulses applied to another quantum dot gate.

\begin{acknowledgments}
We thank Mark Eriksson for useful discussions, Stephan Philips for the design of the PCB onto which the sample was mounted and Francisco Carrasco for assistance with sample fabrication. We acknowledge financial support by Intel Corporation and the QuantERA ERA-NET Cofund in Quantum Technologies implemented within the European Union’s Horizon 2020 Programme.

\end{acknowledgments}

\section*{Data Availability Statement}
The data that supports the findings of this study are available within the article and its supplementary material.

\nocite{*}
\bibliography{mybib}

\end{document}